\documentclass[12pt,a4paper]{article}

\usepackage{amsmath, amssymb, amsthm, bm, amsfonts}
\usepackage{mathtools, booktabs, setspace, natbib}

\usepackage[colorlinks=true,linkcolor=blue, citecolor=blue]{hyperref}
\RequirePackage{graphicx}
\usepackage{caption}
\usepackage{subcaption}
\usepackage{float}
\usepackage[a4paper]{geometry}
\theoremstyle{remark}
\newtheorem{remark}{Remark}

  \def\b1 {{\mathbbm 1}}

\makeatletter
\renewcommand*{\@fnsymbol}[1]{\ifcase#1\or*\else\@arabic{\numexpr#1-1\relax}\fi}
\makeatother

\begin{document}
\title{\bf Practical Forecasting of Environmental Maps: A Functional Data Approach\footnote{
The second author also gratefully acknowledges the financial support from Juan de la Cierva Incorporaci\'{o}n,  IJC2019-041742-I, Ramon y Cajal, RYC2023-044391-I and CEX 2021-001181-M financed by MICIU/AEI /10.13039/501100011033.}}

\author{Alexander Gleim\thanks{Cognite AS, Norway. Email: alexander.gleim@cognite.com} \\ Cognite AS  \and Nazarii Salish\thanks{Corresponding author: Universidad Carlos III de Madrid, Calle Madrid 126, 28903 Getafe (Madrid), Spain. Email: nsalish@eco.uc3m.es} \\ University Carlos III de Madrid}
\maketitle

\begin{abstract}
Environmental problems are receiving increasing attention in socio-economic and health studies, fostering advances in recording and data collection of related real-life processes.
However, traditional tools for data processing are often found too restrictive as they do not account for the rich nature of such data sets.
In this paper, we propose a simple statistical perspective on forecasting environmental data collected sequentially over time across some predefined geographic region.
We treat such data set as a surface (or functional) time series with a possibly complicated geographical domain.
Using techniques from functional data analysis, we develop a forecasting methodology that allows to account for both geographic and temporal dependencies.
This methodology allows integration of traditional multivariate techniques to provide forecasts surfaces.
We demonstrate the practical value of our approach with a forecasting example of ground-level ozone concentration across Germany, showcasing its effectiveness and potential for broad application.
\end{abstract}

%
%

\vspace{1cm}

\newpage
\onehalfspacing

\newpage
\section{Introduction}
\label{sec:introduction}

Environmental data are increasingly used to study health, economic, ecological, and policy-relevant outcomes, and forecasting is one of the main tasks in turning such data into operational information.
Most environmental variables are both geographically distributed and dynamically evolving.
Air pollution, temperature, precipitation, soil moisture, water quality, and solar radiation are observed over geographical regions, often with irregular boundaries, and are recorded repeatedly over time through monitoring stations, or gridded data systems.
When it comes to forecasting such data sets, this creates a forecasting challenge in which the object of interest is not a scalar-valued time series at a single location, but an entire environmental map, typically observed at a set of stations or grid points.
Although the data are observed discretely, the underlying environmental process is naturally distributed over the geographical region of interest.
It is therefore natural to treat such data for what they represent: a time series of surfaces,
\[
    X_t(s), \qquad s \in \mathcal{D}, \quad t=1,\ldots,T,
\]
where \(X_t(s)\) denotes the value of an environmental variable at location \(s\) and time \(t\), and \(\mathcal{D}\) is the geographical domain under study.
The forecasting target is then the future surface \(X_{T+h}(s)\), \(s\in\mathcal{D}\), for horizons \(h\geq 1\).

The purpose of this note is to build a simple and practically useful forecasting procedure for such settings.
By adopting and adjusting recent tools from functional data analysis, we obtain a benchmark method that is easy to implement and competitive in applications.
The approach follows the general logic of functional time-series forecasting: reduce the functional object to a \emph{relevant} low-dimensional vector, forecast this vector with available multivariate tools, and then map the forecast back into the original functional space.
This idea has proved useful for curve time series observed on intervals; see, for instance, \citealp{Hyndman2009} and \citealp{aue2015}.
Here we adapt the same idea to environmental maps observed over geographical domains, but with one important difference: instead of relying on functional principal components, we extract low-dimensional dynamic scores designed to capture the predictable structure of the environmental process.

The proposed workflow has three steps.
First, we reconstruct smooth environmental surfaces and their basis representation from geographically scattered observations using finite-element smoothing, which is convenient for domains with irregular boundaries; see \citealp{Ramsay2002}, \citealp{SangalliRamsayRamsay2013}, and \citealp{Ferraccioli2020}.
Second, we represent the surface time series through a low-dimensional vector that summarizes its predictable dynamic structure.
While FPCA provides a natural starting point for this step and a useful benchmark, its components are selected to explain variation and need not coincide with the directions most relevant for forecasting.
For forecasting environmental maps, the relevant components are those that capture serial dependence and predictable dynamics of the surface process.
We therefore use a dimension-reduction technique developed in \citealp{bathia2010} and \citealp{OttoSalish2026}, where the dynamic component is identified from the cumulative autocovariance structure of the functional time series.
Third, we forecast the resulting low-dimensional vector using multivariate forecasting methods and reconstruct the forecast surface from the predicted values.
Thus, a high-dimensional map-forecasting challenge is transformed into a familiar low-dimensional multivariate forecasting task.

This construction has several practical advantages.
First, it preserves the geographical nature of the data instead of treating monitoring locations as separate scalar series.
Second, it reduces dimensionality and avoids direct estimation of large multivariate models when the number of stations or grid cells is large relative to the length of the time series.
Furthermore, the procedure is modular: once the low-dimensional representation has been obtained, linear, nonlinear, machine-learning, covariate-augmented forecasting methods or other can be used.
Finally, the main components of the workflow are available in standard statistical software such as R and MATLAB; see \citealp{RamsaySilverman2005}.
To improve reproducibility, we collect the three steps of the proposed workflow in a MATLAB package and provide the data used in the empirical illustration at: {\url{https://github.com/nasaza/SurfaceTimeSeries}}.

The paper is related to several strands of literature.
Functional data analysis provides the general framework for representing curves and surfaces as elements of function spaces; see \citealp{RamsaySilverman2005} and \citealp{KokoszkaReimherr2017}.
Functional time-series methods have been developed in, among others, \citealp{Bosq2000}, \citealp{HoermannKokoszka2010}, \citealp{HORMANN2012}, \citealp{salish2019}, and \citealp{Otto2025}.
FPCA-based forecasting is especially relevant as a motivating benchmark.
The approach of \citealp{Hyndman2009} and \citealp{aue2015} shows how functional observations can be summarized by principal-component scores and then forecast using time-series methods.
This FPCA-based approach is also closely related to factor functional representations considered in \citealp{hays2012}, \citealp{Liebl2013}, and \citealp{HORMANN2022a}.
Most of these contributions, however, focus on curves defined on an interval rather than on maps defined over two-dimensional geographical regions.
On the geographical reconstruction side, methods such as kriging, thin-plate splines, soap-film smoothing, and finite-element smoothing have been used to recover continuous surfaces from discrete observations; see, for example, \citealp{Delicadoetal2010}, \citealp{GiraldoDelicadoMateu2009}, \citealp{GiraldoDelicadoMateu2011}, \citealp{NeriniMonestiezMante2010}, \citealp{Oconnell1997}, \citealp{Wood2008}, and \citealp{SangalliRamsayRamsay2013}.
Our contribution is to combine finite-element reconstruction on irregular geographical domains, cumulative autocovariance-based dynamic dimension reduction, and modular multivariate forecasting in a single practical workflow.
By directing dimension reduction towards serial dependence rather than contemporaneous variation, the procedure provides a practical benchmark designed specifically for forecasting environmental maps.

We demonstrate the usefulness of this perspective with an illustration using ground-level ozone concentrations over Germany.
The exercise shows that the workflow produces competitive short-term map forecasts and geographically resolved diagnostics, including MSPE surfaces and visualizations of local exceedance episodes.
This illustration also summarizes the main message of the paper: even without a highly specialized environmental model, a dynamic functional representation of environmental maps provides a useful, reproducible, and practical benchmark for practitioners working with geographically distributed environmental time series.



\section{Forecasting Framework}\label{sec:ForecastingFramework}

In this section, we describe the forecasting framework.
The purpose is to keep the procedure simple: first reconstruct environmental maps as surfaces, then extract a low-dimensional dynamic representation, and finally forecast this representation with standard multivariate tools.

\textbf{STEP 1: Surface Reconstruction and Basis Representation.}
Environmental data are typically collected at discrete grid points or monitoring stations across a geographical domain.
The first step is therefore to transform these discrete observations into surfaces.
Several reconstruction methods could be used for this purpose, including kriging, thin-plate splines, soap-film smoothing, and FEM smoothing.
We use finite-element smoothing because it is well suited to geographical domains with irregular shapes and boundaries; see \citealp{Ramsay2002}, \citealp{SangalliRamsayRamsay2013}, and \citealp{Ferraccioli2020}.
In many environmental applications, the region of interest is not rectangular, and some traditional smoothing methods may smooth across borders or outside the intended domain.
FEM avoids this by reconstructing the surface on a mesh that follows the shape of the region, while remaining computationally simple and efficient.
This step also has a practical computational role.
It not only produces smooth maps but also gives each reconstructed surface a basis representation,
\[
    X_t(s)\approx\sum_{k=1}^{K} b_{k,t}\phi_k(s), \qquad s\in\mathcal D,
\]
where \(\phi_1,\ldots,\phi_K\) are finite-element basis functions and \(b_{1,t},\ldots,b_{K,t}\) are the corresponding coefficients, with $K$ determined by the chosen mesh.
Once the surfaces are represented in this form, inner products, projections and other operations with surface data can be computed through finite-dimensional matrix operations.
This makes the surface time series directly usable in standard statistical software and is essential from an applied implementation point of view.
The technical details of the reconstruction are given in Appendix~\ref{sec:spatial smoothing}.

\emph{Remark on the Grid Density.}
The usefulness of the surface representation depends partly on the density and structure of the observation grid.
When the number of stations or grid cells is large relative to the number of time observations, direct multivariate modelling can become unstable or infeasible.
Moreover, when the grid is sufficiently dense, functional-data estimators often behave as if the underlying functions were fully observed, making the reconstruction step asymptotically negligible for subsequent analysis; see \citealp{hall2006}, \citealp{li2010}, and \citealp{zhang2016}.
This provides an additional motivation for treating dense environmental data as surfaces rather than as unrelated collections of scalar series.

\textbf{STEP 2: Representing Surfaces by Dynamic Scores.}
After reconstructing the surfaces, the next step is to replace each infinite-dimensional observation by a short vector that captures the dynamically relevant part of the surface time series.

A standard approach in the literature is to obtain such a vector through FPCA.
In that case, the directions are chosen to explain the largest share of variation, and the corresponding scores are then forecast with multivariate time-series tools; see \citealp{Hyndman2009} and \citealp{aue2015}.
This provides a convenient benchmark that is extensively used in applied research.
For forecasting, however, directions that explain the most variation need not be the directions that are most useful for prediction.
A surface component may be highly variable but weakly predictable, while another component may explain less variation but carry more serial dependence.
We therefore use an autocovariance-based dimension-reduction step, following the ideas of \citealp{bathia2010} and \citealp{OttoSalish2026}.
The aim is to extract directions that capture serial dependence and predictable dynamics rather than contemporaneous variation.

In simple terms, the process can be decomposed into three parts:
\begin{equation*}
  X_t(s)  =    \mu(s)+\sum_{\ell=1}^{L} z_{\ell,t}\psi_\ell(s) + \varepsilon_t(s),
    \qquad s\in\mathcal D,
\end{equation*}
where \(\mu\) is the mean surface, \(\psi_1,\ldots,\psi_L\) are deterministic dynamic directions, $z_{\ell,t}=\langle X_t-\mu,\psi_\ell\rangle$ with $\ell=1,\ldots,L,$ are the corresponding dynamic scores and $\varepsilon_t(s)$ denotes the remainder outside the retained dynamic subspace.
Then the vector
\[
   \mathbf z_t=(z_{1,t},\ldots,z_{L,t})'
\]
is the low-dimensional representation of the dynamic core of the original surface \(X_t\).
Thus, the form of the representation is similar to FPCA, but crucially the directions, $\psi_\ell(s)$, are identified (and estimated) differently: FPCA uses contemporaneous variation, whereas the present approach uses autocovariances to recover the part of the surface time series that is relevant for prediction.
A formal description of this decomposition and its estimation is provided in Appendix~\ref{app:fpca}.
The number of retained components \(L\) can be selected using the information criteria proposed by \citet{OttoSalish2026}.

\textbf{STEP 3: Forecasting and Reconstruction.}
The final objective is to obtain the \(h\)-step-ahead forecast of the surface \(X_{t+h}\), given past surfaces and any additional covariates.
After Steps 1 and 2, the problem reduces to forecasting the multivariate time series of dynamic scores \(\mathbf  z_t\).
Let \(\mathbf  W_t\) denote additional information available at time \(t\), such as meteorological covariates.
A general score forecast can be written as
\[
    \widehat{\mathbf{z}}_{t+h|t}
    =
    \mathcal F_h(\mathbf z_t,\mathbf z_{t-1},\ldots,\mathbf W_t),
\]
where \(\mathcal F_h\) may be linear or nonlinear. Once \(\widehat{\mathbf{z}}_{t+h|t}\) is obtained, the forecast surface is reconstructed as
\[
    \widehat X_{t+h|t}(s)
    =
    \widehat\mu(s)
    +
    \sum_{\ell=1}^{L}
    \widehat z_{\ell,t+h|t}\widehat\psi_\ell(s),
    \qquad s\in\mathcal D.
\]
Thus, all forecasting is carried out in a low-dimensional vector space, while the final output remains a full environmental map.
The main advantage of the framework is its modularity: while steps 1 and 2 remain the same, the forecasting method can be adapted to the data, the forecast horizon, and the analyst’s goal (e.g., linear, non-linear, machine-learning methods etc).
This also leaves room for future extensions, since more advanced or newly developed forecasting tools can be incorporated without changing the remaining parts of the pipeline.

\emph{Remark: A multivariate route to the scores.}
The previous steps first reconstruct surfaces and then estimate their low-dimensional representation.
As an alternative, one can estimate the score vectors directly from the observed grid or station data, following the multivariate perspective developed in \citealp{HORMANN2022a} and \citealp{HORMANN2022b}.
This can simplify the estimation of the score vectors when the data are observed on a sufficiently rich grid.
However, even if the scores are estimated directly from the discrete observations, the final forecast still has to be mapped back into a surface.
For this reason, estimates of the mean surface and the relevant directions are still needed for producing forecast maps.
Technical details are collected in Appendix~\ref{sec:MultivPersp}.

\section{An Empirical Illustration: Ozone Surfaces}\label{sec:Results_SFDA}

We illustrate the proposed framework with daily ground-level ozone concentrations over Germany.
Ground-level ozone is a harmful secondary pollutant formed through photochemical reactions involving precursor pollutants in the presence of sunlight.
Because its effects on respiratory health, vegetation, and crop yields are well documented (see, e.g., \citealp{Stewart2017}), geographically resolved forecasts are useful for monitoring, early warnings, and the identification of local areas where air-quality thresholds may be exceeded.
The purpose of this illustration is not to conduct an exhaustive comparison with the full air-quality forecasting literature, which often relies on machine-learning methods for station-level prediction (see, e.g., \citealp{FENG2011}, \citealp{GONG2016}, and \citealp{MARVIN2022}).
Instead, the aim is to show how a surface-based statistical benchmark can complement existing tools by producing forecasts for the full geographical map rather than only for a set of individual stations.

\textbf{Data and preprocessing.}
The ozone data are daily measurements from AirBase, the European air-quality database provided by the European Environment Agency.
We use the 171 German monitoring stations with complete daily observations throughout 2011, measured in \(\mu\mathrm{g}/\mathrm{m}^3\).
As additional covariates, we use temperature, precipitation, solar radiation, and wind speed from the E-OBS data set, available on a regular \(0.25^\circ\times0.25^\circ\) latitude--longitude grid.
All variables display strong seasonal patterns, which we remove by subtracting monthly averages at each location.
The cleaned data, seasonal components, metadata, and scripts used to reproduce the preprocessing steps are provided in the supplementary MATLAB package.
Further details on data construction and preprocessing are provided in Appendix~\ref{app:data_preprocessing}.

For each day, ozone measurements are reconstructed as a surface over the German domain using finite-element smoothing.
The same reconstruction procedure is applied to the meteorological covariates.
The dynamic low-dimensional representation is then estimated from the cumulative autocovariance structure of the ozone surface time series, as described in Section~\ref{sec:ForecastingFramework}.
The information criterion of \citet{OttoSalish2026} selects four dynamic scores, whereas the functional final prediction error criterion of
\citet{aue2015} selects eight FPCA scores.
Both criteria select a lag order of two for the corresponding forecasting specifications.
Thus, in this application, the cumulative-autocovariance approach provides a substantially more parsimonious representation of the forecast-relevant dynamics.
Appendix Figure~\ref{fig:PCA_DS_directions} compares the leading FPCA and dynamic directions, illustrating how variance-based and cumulative-autocovariance-based dimension reduction emphasize different geographical patterns.

The data contain several local exceedance episodes.
In particular, there are 28 days in 2011 on which ozone concentration exceeds \(100~\mu\mathrm{g}/\mathrm{m}^3\) in at least one part of the domain.
The largest observed value is \(123.74~\mu\mathrm{g}/\mathrm{m}^3\), recorded on 31 May near Dresden.
These episodes are useful for evaluating whether the forecasting framework can identify localized high-ozone events, not only average map accuracy.

\textbf{Forecasting design and benchmarks.}
The first 200 days are used as the initial training sample, while the remaining 165 days are used for forecast evaluation.
We use a rolling-origin scheme: after each forecast, the estimation sample is expanded by one day and the models are re-estimated.
We report one-step-ahead results in the main text, while the corresponding boxplots and MSPE surfaces for the 3-step- and 7-step-ahead forecasts are reported in Appendix~\ref{sec:AdditionalResults}.

The comparison includes several simple benchmarks and alternative implementations of the proposed workflow.
The mean forecast (MF) uses the historical average surface, while the naive forecast (NF) uses the last observed surface.
The functional autoregressive forecast (FAR) is included as a standard benchmark for functional time series; see, for example, \citealp{didericksen2012}.
We also compare FPCA-based score forecasts using a covariate-augmented VAR and a \(k\)-nearest-neighbour method (PCA\_VAR and PCA\_KNN) with dynamic-score forecasts based on the same forecasting tools (DS\_VAR and DS\_KNN).
To illustrate the multivariate route discussed in Section~\ref{sec:ForecastingFramework}, we also estimate score vectors directly from the station data and forecast them by VAR and KNN (MP\_VAR and MP\_KNN).
Finally, we include a random-forest implementation applied to the dynamic scores (DS\_RF).
Implementation details for the forecasting methods are provided in Appendix~\ref{sec:Forecasting}.

\textbf{Forecast evaluation.}
Let \(\mathcal T_h\) denote the set of forecast origins for horizon \(h\), and let \(s_1,\ldots,s_N\) be the station locations used for evaluation.
We summarize geographical forecast accuracy by the mean squared prediction error surface
\[
    \mathrm{MSPE}_{h}(s_i)
    =
    \frac{1}{|\mathcal T_h|}
    \sum_{t\in\mathcal T_h}
    \left\{X_{t+h}(s_i)-\widehat X_{t+h|t}(s_i)\right\}^2,
    \qquad i=1,\ldots,N .
\]
We also report the spatially averaged error $ \overline{\mathrm{MSPE}}_{h} = \frac{1}{N}   \sum_{i=1}^{N}\mathrm{MSPE}_{h}(s_i), $
for each model and horizon.
This pair of summaries is useful: \(\overline{\mathrm{MSPE}}_{h}\) compares models on average, while \(\mathrm{MSPE}_{h}(s_i)\) shows where a method performs well or poorly across the map of Germany.

\begin{figure}[tp]
\centering
  \includegraphics[width=0.85\textwidth]{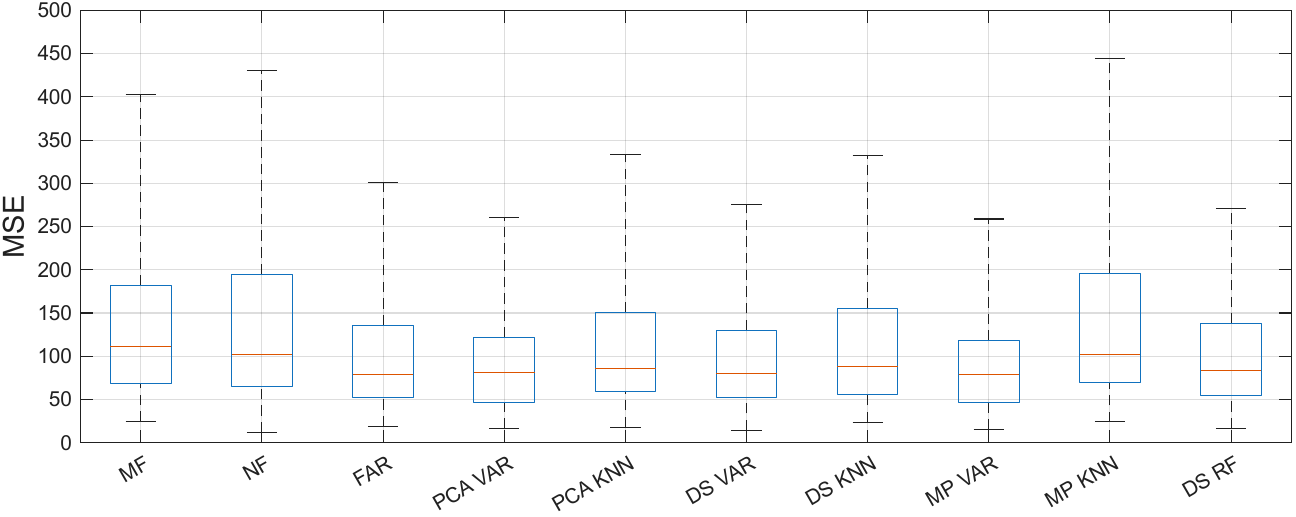}
  \caption{{\footnotesize \textbf{Boxplots of one-step-ahead forecast errors.}
  The figure reports rolling-origin forecast errors for the benchmark and surface-time-series methods.
  MF: mean forecast; NF: naive forecast; FAR: functional autoregression; PCA\_VAR/PCA\_KNN: FPCA-score forecasts; DS\_VAR/DS\_KNN: dynamic-score forecasts; MP\_VAR/MP\_KNN: multivariate-perspective forecasts; DS\_RF: random forest on dynamic scores.}}
  \label{fig:Forecasts1}
\end{figure}

Across the three forecast horizons, the linear specifications PCA\_VAR, DS\_VAR, and MP\_VAR provide the best overall forecasting performance according to the geographically averaged MSPE; see Figures~\ref{fig:Forecasts1}, \ref{fig:Forecasts3}, and \ref{fig:Forecasts7}. Including meteorological covariates improves the performance of all three specifications. To conserve space, we do not report the corresponding results without covariates; excluding them increases forecast errors but does not alter the overall ranking, with the linear specifications remaining the strongest performers.

\begin{figure}[tp]
\centering
  \includegraphics[width=0.85\textwidth]{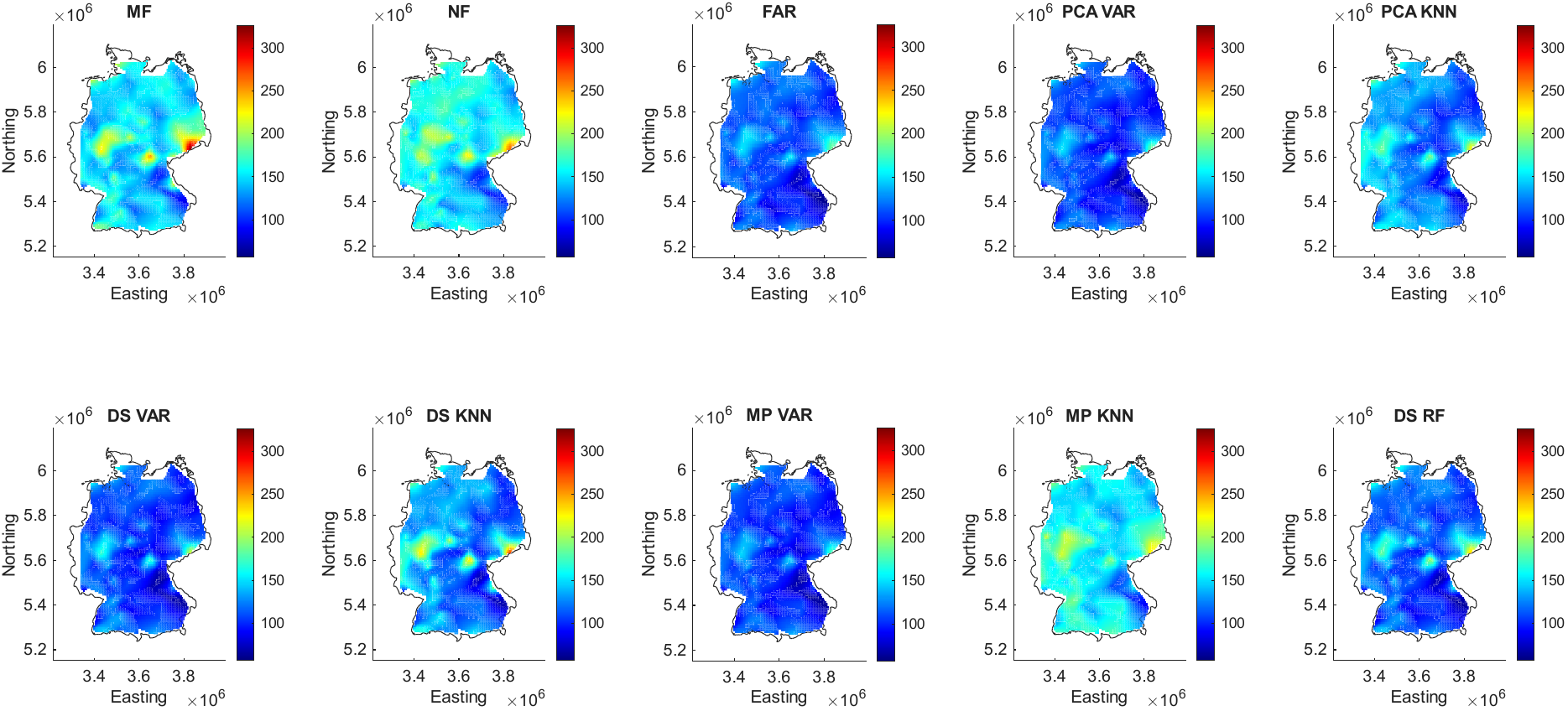}
  \caption{{\footnotesize \textbf{MSPE surfaces for one-step-ahead forecasts.}
  Each panel shows \(\mathrm{MSPE}_{1}(s)\) over the reconstruction domain.
  The panels reveal where each forecasting method performs well or poorly across Germany.}}
  \label{fig:MSESurfaces1}
\end{figure}

Surface forecasts also make it possible to inspect errors geographically.
Figure~\ref{fig:MSESurfaces1} reports the MSPE surfaces for the one-step-ahead forecasts, while Figures~\ref{fig:MSESurfaces3} and \ref{fig:MSESurfaces7} report the corresponding results for longer horizons.
These diagnostics complement the average MSPE by showing how forecast accuracy varies across Germany.
The MSPE surfaces reinforce the aggregate results: the covariate-augmented linear models generally produce smaller forecast errors across most of the country.
All methods nevertheless exhibit persistent error ``hotspots'' in the south-west and in parts of central and eastern Germany.
These patterns suggest that further gains may come from incorporating additional local information, such as terrain, urban density, industrial activity, or other environmental variables, and/or improving monitoring-station coverage.

\begin{figure}[tp]
\centering
  \includegraphics[width=0.85\textwidth]{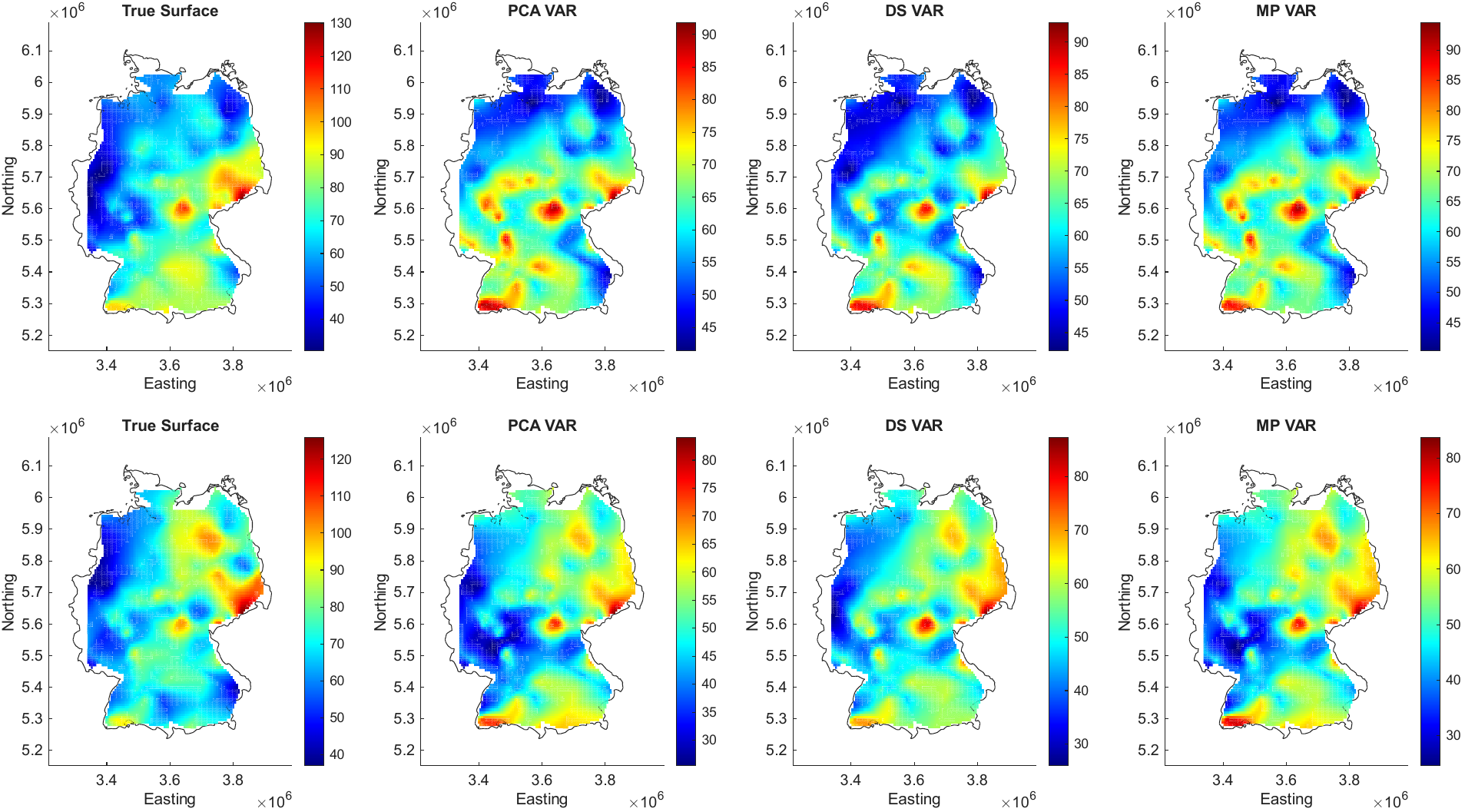}
  \caption{{\footnotesize \textbf{Observed and forecast ozone surfaces during local exceedance episodes.}
  Top row: 24 August; bottom row: 26 August.
  The figure compares the observed surface with selected forecast surfaces and highlights whether the methods anticipate the local high-ozone regions.}}
   \label{fig:SpecialEvents}
\end{figure}

We also use the surface forecasts to examine localized exceedance episodes.
The hold-out period contains two exceedance events around Dresden on 24 and 26 August.
Figure~\ref{fig:SpecialEvents} compares the observed ozone surfaces with selected forecasts for these dates.
The figure therefore assesses not only numerical accuracy but also whether the methods correctly locate the high-ozone region.
The three linear specifications shown in Figure~\ref{fig:SpecialEvents} identify the high-ozone region around Dresden on both dates.
However, these models overpredict ozone concentrations in south-western Germany during these episodes---a region that also displays relatively high MSPEs in general (see Figure~\ref{fig:MSESurfaces1}).

The main lesson from the illustration is that the proposed workflow turns irregularly observed environmental measurements into operational forecast maps.
It provides standard accuracy summaries, geographically resolved error diagnostics, and visual tools for local threshold events within one coherent and reproducible pipeline.

\singlespacing
\bibliographystyle{chicago}
\bibliography{literature_SFDA.bib}

\newpage

\appendix

\section{Forecasting Methodology}
\label{app:forecasting_methodology}
\setcounter{equation}{0} \renewcommand{\theequation}{A.\arabic{equation}}
\setcounter{figure}{0} \renewcommand{\thefigure}{A.\arabic{figure}}
\setcounter{table}{0} \renewcommand{\thetable}{A.\arabic{table}}

\subsection{Surface Reconstruction}
\label{sec:spatial smoothing}

In practice, the surface \(X_t\) is observed, possibly with measurement error, only at discrete locations
\(\{s_i\}_{i=1}^{N}\subset\mathcal D\).
A smoothing procedure is therefore required to reconstruct \(X_t(s)\) for every
\(s\in\mathcal D\) and \(t=1,\ldots,T\).
We use a finite-element spline smoother; see \citealp{Ramsay2002},
\citealp{SangalliRamsayRamsay2013}, and \citealp{Ferraccioli2020}.
The reconstruction proceeds in three steps.

\begin{description}

\item[\textbf{Step 1.}]
\textbf{Triangulate the geographical domain.}
The domain \(\mathcal D\) is partitioned into non-overlapping triangles whose vertices include the observation locations.
We use a Delaunay triangulation, which avoids unnecessarily thin triangles and favours triangles that are close to equiangular.
The resulting triangulation is denoted by \(\triangle_{\mathcal D}\).
Figure~\ref{fig:stations+mesh} illustrates the mesh used for the German ozone-monitoring network.

\begin{figure}[th]
\centering
  \includegraphics[width=0.4\textwidth]{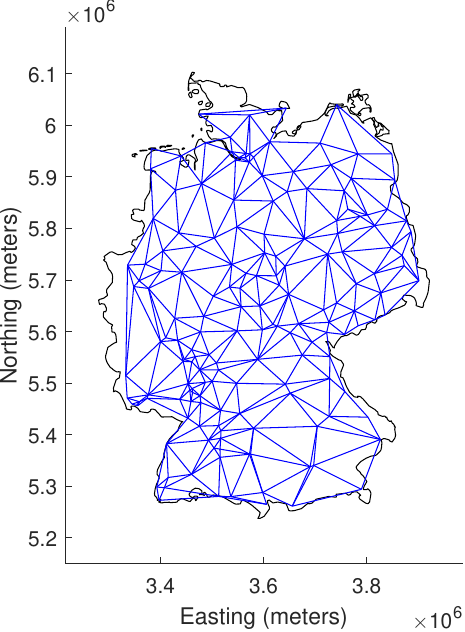}
  \caption{{\footnotesize \textbf{Triangular mesh.}
  Delaunay triangulation with nodes located at the ozone-monitoring stations.}}
  \label{fig:stations+mesh}
\end{figure}

\item[\textbf{Step 2.}]
\textbf{Construct the finite-element basis.}
On every triangle, the surface is represented by a quadratic polynomial that is continuous across common edges and vertices.
A quadratic triangular element is determined by six local nodes: the three vertices and the midpoint of each edge.
Associated with these nodes are six quadratic shape functions, each equal to one at its own node and zero at the remaining local nodes; see Figure~\ref{fig:QFE shape functions}.

\begin{figure}[th]
\centering
  \includegraphics[width=0.9\textwidth]{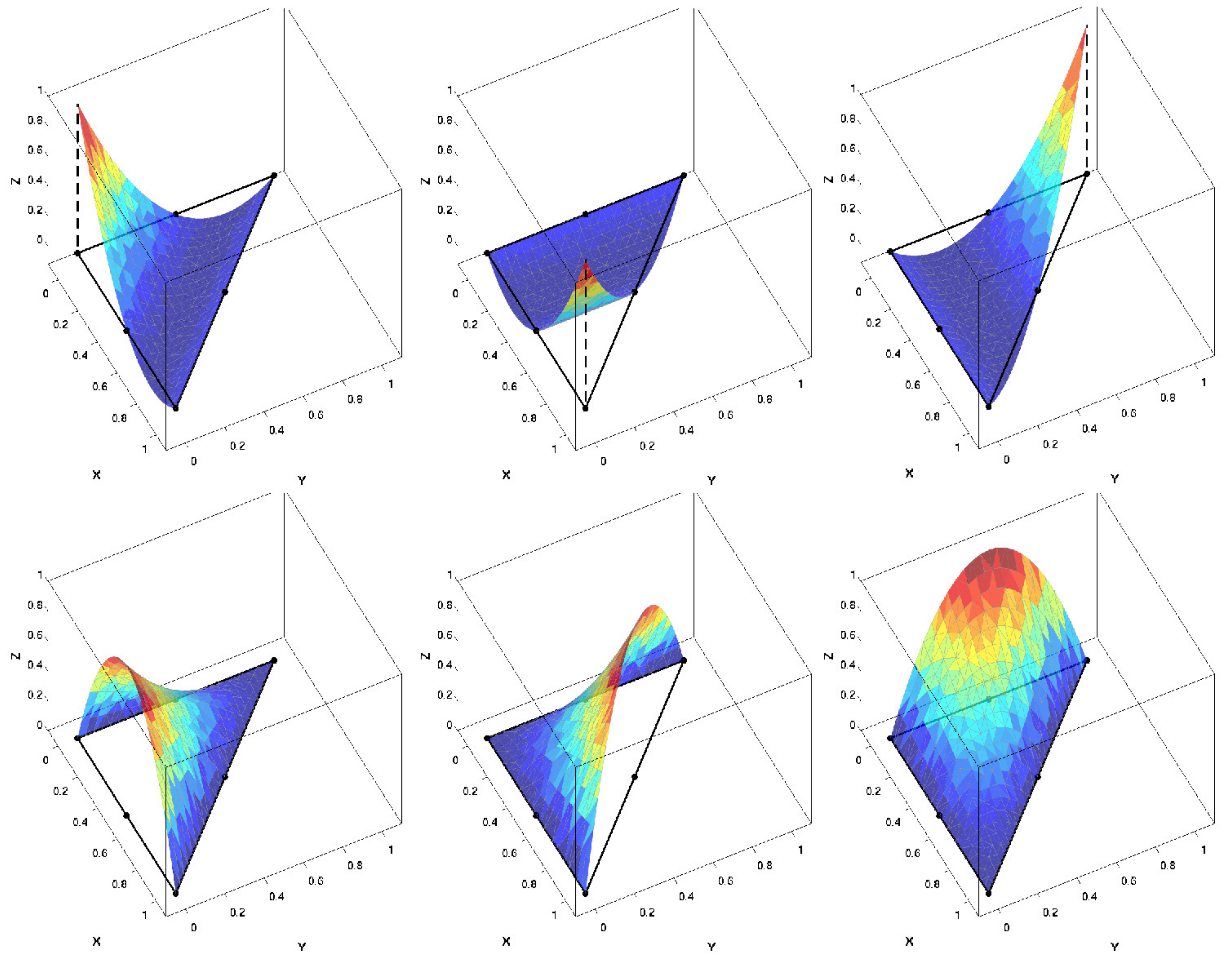}
  \caption{{\footnotesize \textbf{Quadratic finite-element shape functions.}
  The six shape functions on a triangular reference element.}}
  \label{fig:QFE shape functions}
\end{figure}

Let \(\xi_1,\ldots,\xi_K\) denote the union of all nodal points in the triangulation.
The nodes are numbered so that the observation locations \(s_1,\ldots,s_N\) correspond to the first \(N\) nodes.
With every node \(\xi_k\), we associate a global nodal basis function \(\phi_k\).
The collection \(\{\phi_k\}_{k=1}^{K}\) spans the space of continuous functions on
\(\mathcal D\) that are quadratic polynomials on each triangle.

\item[\textbf{Step 3.}]
\textbf{Obtain the basis coefficients.}
Let
\[
    \boldsymbol{\phi}(s)
    =
    \bigl(\phi_1(s),\ldots,\phi_K(s)\bigr)^\prime .
\]
The reconstructed surface has the finite representation
\[
    X_t(s)
    \approx
    \sum_{k=1}^{K} b_{k,t}\phi_k(s)
    =
    \mathbf b_t^\prime\boldsymbol{\phi}(s),
    \qquad t=1,\ldots,T.
\]
Define the finite-element mass and stiffness matrices
\[
    \mathbf G
    =
    \int_{\mathcal D}
    \boldsymbol{\phi}(s)\boldsymbol{\phi}(s)^\prime\,ds
\]
and
\[
    \mathbf P
    =
    \int_{\mathcal D}
    \left[
    \boldsymbol{\phi}^{(x)}(s)\boldsymbol{\phi}^{(x)}(s)^\prime
    +
    \boldsymbol{\phi}^{(y)}(s)\boldsymbol{\phi}^{(y)}(s)^\prime
    \right]ds,
\]
where \(\boldsymbol{\phi}^{(x)}\) and \(\boldsymbol{\phi}^{(y)}\) collect the partial derivatives of the basis functions.
Let \(\mathbf D\) be the \(K\times K\) diagonal matrix whose first \(N\) diagonal elements equal one and whose remaining elements equal zero.
Finally, let
\[
    \mathbf X_t^\ast
    =
    \bigl(
    X_t(s_1),\ldots,X_t(s_N),0,\ldots,0
    \bigr)^\prime .
\]
Following \citet{SangalliRamsayRamsay2013}, the coefficient vector \(\mathbf b_t\) is obtained by solving
\begin{equation*}
  \begin{pmatrix}
    -\mathbf D & \lambda\mathbf P\\
    \lambda\mathbf P & \lambda\mathbf G
  \end{pmatrix}
  \begin{pmatrix}
    \mathbf b_t\\
    \mathbf v_t
  \end{pmatrix}
  =
  \begin{pmatrix}
    -\mathbf X_t^\ast\\
    \mathbf 0
  \end{pmatrix},
\end{equation*}
where \(\lambda>0\) controls the degree of smoothing and \(\mathbf v_t\) is an auxiliary vector.
The same mesh, basis, and smoothing parameter are used for all observations in a given surface time series.

\end{description}

\begin{remark}[Choice of the smoothing parameter]
\label{rem:gcv}
A common data-driven choice of \(\lambda\) minimizes the generalized cross-validation criterion
\begin{equation*}
  \operatorname{GCV}(\lambda)
  =
  \frac{
  \bigl(\mathbf X_N^\ast-\mathbf S_N(\lambda)\mathbf X_N^\ast\bigr)^\prime
  \bigl(\mathbf X_N^\ast-\mathbf S_N(\lambda)\mathbf X_N^\ast\bigr)}
  {N\left\{1-\operatorname{tr}\bigl(\mathbf S_N(\lambda)\bigr)/N\right\}^2},
\end{equation*}
where \(\mathbf S_N(\lambda)\) is the \(N\times N\) block of the smoothing matrix that maps observations at the monitoring locations into fitted values at those locations.
See \citet{RamsaySilverman2005} for further discussion.
\end{remark}

The basis representation is also the computational input for the remaining steps.
For two reconstructed surfaces with coefficient vectors \(\mathbf a\) and \(\mathbf b\),
\[
    \langle a,b\rangle
    =
    \mathbf a^\prime\mathbf G\mathbf b.
\]
Thus, functional inner products, projections, and operator estimators can be evaluated by finite-dimensional matrix operations.

\subsection{Dynamic Scores}
\label{app:fpca}

Let \(\mathcal H=L^2(\mathcal D)\) be the Hilbert space of square-integrable surfaces on
\(\mathcal D\), equipped with inner product
\[
    \langle f,g\rangle
    =
    \int_{\mathcal D}f(s)g(s)\,ds
\]
and norm \(\|f\|^2=\langle f,f\rangle\).
Let \(\mu(s)=E[X_t(s)]\) and define the centered surface
\(Y_t=X_t-\mu\).

\subsubsection*{Cumulative autocovariance operator representation}

A standard FPCA representation chooses directions from the covariance operator and therefore orders them by contemporaneous variation.
For forecasting, however, a direction may explain substantial variation while containing little information about future surfaces.
Following \citet{bathia2010} and \citet{OttoSalish2026}, we instead identify the finite-dimensional dynamic subspace from the autocovariance structure of \(\{X_t\}\).

For an integer lag \(\tau\geq1\), define the lag-\(\tau\) autocovariance operator
\[
    C_\tau f
    =
    E\!\left[
       Y_t\langle Y_{t-\tau},f\rangle
    \right],
    \qquad f\in\mathcal H.
\]
Its adjoint is denoted by \(C_\tau^\ast\).
For a fixed, typically small, integer \(q\), define the cumulative autocovariance operator
\begin{equation*}
    \mathcal K_q
    =
    \sum_{\tau=1}^{q} C_\tau C_\tau^\ast .
\end{equation*}
The operator \(\mathcal K_q\) is self-adjoint, positive semidefinite, and compact.
Let
\[
    \nu_1\geq\cdots\geq\nu_L>0
\]
be its non-zero eigenvalues and
\(\psi_1,\ldots,\psi_L\) the corresponding orthonormal eigenfunctions:
\[
    \mathcal K_q\psi_\ell
    =
    \nu_\ell\psi_\ell,
    \qquad \ell=1,\ldots,L.
\]

Under the conditions in \citet{bathia2010} and \citet{OttoSalish2026}, the range of
\(\mathcal K_q\),
\[
    \mathcal H_D
    =
    \operatorname{span}(\psi_1,\ldots,\psi_L),
\]
is the dynamic subspace of the process.
It contains all directions along which the surface time series exhibits serial dependence.
The dimension \(L=\operatorname{rank}(\mathcal K_q)\) is therefore the dimension of the dynamically relevant part of the process, rather than the number of directions required to explain a chosen percentage of contemporaneous variance.

The corresponding dynamic scores are
\begin{equation*}
    z_{\ell,t}
    =
    \langle X_t-\mu,\psi_\ell\rangle,
    \qquad \ell=1,\ldots,L,
\end{equation*}
and we write
\[
    \mathbf z_t
    =
    (z_{1,t},\ldots,z_{L,t})^\prime .
\]
The surface process then admits the decomposition
\begin{equation}
\label{eq:dynamic_decomposition_appendix}
    X_t(s)
    =
    \mu(s)
    +
    \sum_{\ell=1}^{L}z_{\ell,t}\psi_\ell(s)
    +
    \varepsilon_t(s).
\end{equation}
The first two terms form the dynamically relevant component of \(X_t\).
Under the assumptions used in \citet{OttoSalish2026}, the remainder
\(\varepsilon_t\) carries no additional predictable information from the past of the surface process.
Equation~\eqref{eq:dynamic_decomposition_appendix} has the same convenient form as a truncated FPCA representation, but the directions are selected using autocovariances rather than the covariance operator.
This aligns the dimension-reduction step directly with the forecasting objective.

\subsubsection*{Estimation}

The population quantities are replaced by their sample analogues.
The mean surface is estimated by
\[
    \widehat\mu(s)
    =
    \frac{1}{T}\sum_{t=1}^{T}X_t(s),
\]
and the centered observations are
\(\widehat Y_t=X_t-\widehat\mu\).
For \(\tau=1,\ldots,q\), define
\[
    \widehat C_\tau f
    =
    \frac{1}{T-\tau}
    \sum_{t=\tau+1}^{T}
    \widehat Y_t
    \langle\widehat Y_{t-\tau},f\rangle .
\]
The empirical cumulative autocovariance operator is
\begin{equation}
\label{eq:estimated_dynamic_operator}
    \widehat{\mathcal K}_q
    =
    \sum_{\tau=1}^{q}
    \widehat C_\tau\widehat C_\tau^\ast .
\end{equation}
Let
\(\widehat\nu_1\geq\widehat\nu_2\geq\cdots\geq0\)
and
\(\widehat\psi_1,\widehat\psi_2,\ldots\)
be its eigenvalues and orthonormal eigenfunctions.
For a selected dimension \(L\), the estimated scores are
\begin{equation*}
    \widehat z_{\ell,t}
    =
    \langle X_t-\widehat\mu,\widehat\psi_\ell\rangle,
    \qquad
    \ell=1,\ldots,L.
\end{equation*}
The estimated dynamically relevant surface is
\[
    \widehat X_t^{(L)}(s)
    =
    \widehat\mu(s)
    +
    \sum_{\ell=1}^{L}
    \widehat z_{\ell,t}\widehat\psi_\ell(s).
\]

The lag cutoff \(q\) is used only to identify the dynamic subspace and need not equal the lag order of the forecasting model.
In applications, a small value of \(q\) is often sufficient.
The dimension \(L\) can be selected using an eigenvalue-ratio rule based on \(\widehat{\mathcal K}_q\), or jointly with the forecasting lag order using the information criteria proposed by \citet{OttoSalish2026}.

\subsubsection*{Computation in the finite-element basis}

The finite-element representation from Appendix~\ref{sec:spatial smoothing} makes the estimator in
\eqref{eq:estimated_dynamic_operator} directly computable.
Write
\[
    X_t(s)
    =
    \mathbf b_t^\prime\boldsymbol{\phi}(s),
    \qquad
    \widehat\mu(s)
    =
    \overline{\mathbf b}^{\,\prime}\boldsymbol{\phi}(s),
\]
where
\[
    \overline{\mathbf b}
    =
    \frac{1}{T}\sum_{t=1}^{T}\mathbf b_t,
    \qquad
    \widetilde{\mathbf b}_t
    =
    \mathbf b_t-\overline{\mathbf b}.
\]
Recall that \(\mathbf G\) is the finite-element mass matrix, so that
\[
    \langle X_t-\widehat\mu,X_j-\widehat\mu\rangle
    =
    \widetilde{\mathbf b}_t^\prime
    \mathbf G
    \widetilde{\mathbf b}_j.
\]

Let \(\mathbf R\) be an upper-triangular Cholesky factor satisfying
\[
    \mathbf G=\mathbf R^\prime\mathbf R,
\]
and define the transformed coefficient vectors
\[
    \mathbf u_t
    =
    \mathbf R\widetilde{\mathbf b}_t .
\]
These vectors are expressed in Euclidean coordinates that preserve the functional inner product.
For each lag \(\tau\), form
\[
    \widehat{\boldsymbol\Gamma}_\tau
    =
    \frac{1}{T-\tau}
    \sum_{t=\tau+1}^{T}
    \mathbf u_t\mathbf u_{t-\tau}^\prime
\]
and then
\begin{equation*}
    \widehat{\mathbf M}_q
    =
    \sum_{\tau=1}^{q}
    \widehat{\boldsymbol\Gamma}_\tau
    \widehat{\boldsymbol\Gamma}_\tau^\prime .
\end{equation*}
Let \(\widehat{\mathbf v}_\ell\) be an orthonormal eigenvector of
\(\widehat{\mathbf M}_q\).
The coefficient vector of the corresponding estimated dynamic direction is
\[
    \widehat{\boldsymbol\theta}_\ell
    =
    \mathbf R^{-1}\widehat{\mathbf v}_\ell,
\]
so that
\[
    \widehat\psi_\ell(s)
    =
    \widehat{\boldsymbol\theta}_\ell^\prime
    \boldsymbol{\phi}(s).
\]
The estimated score can be computed equivalently as
\begin{equation*}
    \widehat z_{\ell,t}
    =
    \widetilde{\mathbf b}_t^\prime
    \mathbf G
    \widehat{\boldsymbol\theta}_\ell
    =
    \mathbf u_t^\prime\widehat{\mathbf v}_\ell .
\end{equation*}

\subsection{Forecasting}
\label{sec:Forecasting}

Let
\[
    \boldsymbol\Psi(s)
    =
    \bigl(\psi_1(s),\ldots,\psi_L(s)\bigr)^\prime
\]
and let \(\mathbf z_t=(z_{1,t},\ldots,z_{L,t})^\prime\) denote the dynamic-score vector.
Under decomposition~\eqref{eq:dynamic_decomposition_appendix}, the dynamically relevant part of the surface is
\[
    X_t^{D}(s)
    =
    \mu(s)+\boldsymbol\Psi(s)^\prime\mathbf z_t.
\]
Let \(\mathcal I_t\) denote the information available at time \(t\), including past surfaces and any additional covariates.
The \(h\)-step score predictor is
\begin{equation*}
    \mathbf z_{t+h|t}
    =
    E(\mathbf z_{t+h}\mid\mathcal I_t).
\end{equation*}
Because the remainder in
\eqref{eq:dynamic_decomposition_appendix}
contains no additional predictable component under the maintained assumptions, the corresponding surface predictor is
\begin{equation}
\label{eq:forecastSurface}
    X_{t+h|t}(s)
    =
    \mu(s)
    +
    \boldsymbol\Psi(s)^\prime\mathbf z_{t+h|t}.
\end{equation}
In practice, all unknown quantities in
\eqref{eq:forecastSurface}
are replaced by estimates.

\subsubsection{Linear Forecast: VARX}
\label{sec:LinFor}

A convenient linear model for the score vector is a vector autoregression augmented with exogenous covariates:
\begin{equation}
\label{eq:forecLin}
    \mathbf z_t
    =
    \mathbf c
    +
    \sum_{j=1}^{p}\mathbf A_j\mathbf z_{t-j}
    +
    \sum_{r=0}^{q_w}\mathbf B_r\mathbf w_{t-r}
    +
    \mathbf u_t,
\end{equation}
where
\(\mathbf A_j\) is \(L\times L\),
\(\mathbf w_t\) is an \(M\times1\) vector of covariates,
and
\(\mathbf B_r\) is \(L\times M\).
The covariate vector may contain directly observed scalar variables or low-dimensional scores extracted from surface-valued covariates.

Given information available at time \(t\), the one-step forecast is
\[
    \widehat{\mathbf z}_{t+1|t}
    =
    \widehat{\mathbf c}
    +
    \sum_{j=1}^{p}
    \widehat{\mathbf A}_j\widehat{\mathbf z}_{t+1-j}
    +
    \sum_{r=0}^{q_w}
    \widehat{\mathbf B}_r\mathbf w_{t+1-r|t},
\]
where \(\mathbf w_{t+1-r|t}\) is observed when available at the forecast origin and otherwise replaced by its own forecast.
Multi-step forecasts are obtained recursively.

The implementation has three steps:
\begin{description}
\item[\textbf{Step 1.}]
Estimate the dynamic directions and score vectors using Appendix~\ref{app:fpca}.
If a covariate is surface-valued, construct its low-dimensional representation separately.

\item[\textbf{Step 2.}]
Choose the dimension \(L\), the VAR order \(p\), and, when relevant, the cumulative autocovariance order \(q_w\).
For the dynamic-score representation, \(L\) and \(p\) may be selected jointly using the information criteria of \citet{OttoSalish2026}.
When dynamic scores are replaced by FPCA scores, the number of retained principal components and \(p\) may instead be selected jointly using the
functional final prediction error criterion of \citet{aue2015}.

\item[\textbf{Step 3.}]
Estimate model~\eqref{eq:forecLin}, obtain
\(\widehat{\mathbf z}_{t+h|t}\), and reconstruct the forecast surface using
\eqref{eq:forecastSurface}.
\end{description}

\subsubsection{Nonlinear Forecast: \(k\)-Nearest Neighbours}
\label{sec:nonLinFor}

The KNN method provides a simple nonlinear alternative.
Let
\[
    \mathbf r_t
    =
    \bigl(
    \mathbf z_t^\prime,
    \mathbf z_{t-1}^\prime,
    \ldots,
    \mathbf z_{t-p+1}^\prime
    \bigr)^\prime
\]
be the current state vector.
The forecasting relation is written as
\begin{equation*}
    \mathbf z_{t+h|t}
    =
    m_h(\mathbf r_t),
\end{equation*}
where \(m_h(\cdot)\) is estimated locally.
This approach applies standard multivariate KNN forecasting after the functional dimension-reduction step; see \citealp{Yakowitz1987}.
Related KNN methods developed directly for functional data include
\citealp{Biau2010}, \citealp{KUDRASZOW2013}, and \citealp{KARA2017}.

For every admissible historical state \(\mathbf r_s\), compute its distance from \(\mathbf r_t\), and let
\(\mathcal I_{\widehat K}(t)\) contain the indices of the
\(\widehat K\) nearest neighbours.
The \(h\)-step forecast is
\begin{equation*}
    \widehat{\mathbf z}_{t+h|t}
    =
    \sum_{s\in\mathcal I_{\widehat K}(t)}
    \omega_{s,t}\mathbf z_{s+h},
    \qquad
    \sum_{s\in\mathcal I_{\widehat K}(t)}
    \omega_{s,t}=1.
\end{equation*}
Equal weights set
\(\omega_{s,t}=1/\widehat K\).
Alternatively, normalized inverse-distance weights may be used.
The number of neighbours \(\widehat K\), the state length \(p\), and any distance or weighting parameters are selected by cross-validation within the training sample.
Only historical states for which the subsequent value \(\mathbf z_{s+h}\) is observed are eligible.
Finally, the forecast surface is reconstructed using
\eqref{eq:forecastSurface}.

\subsubsection{Multivariate Perspective}
\label{sec:MultivPersp}

As an alternative benchmark, the low-dimensional score vectors can be estimated directly from the observations at the grid or station locations, without first reconstructing every surface.
This follows the multivariate perspective developed in
\citealp{HORMANN2022a} and \citealp{HORMANN2022b}.

Let
\[
    \mathbf y_t
    =
    \bigl(
    X_t(s_1),\ldots,X_t(s_N)
    \bigr)^\prime
\]
and let
\[
    \mathbf Y
    =
    \bigl(
    \mathbf y_1-\overline{\mathbf y},
    \ldots,
    \mathbf y_T-\overline{\mathbf y}
    \bigr)
\]
be the \(N\times T\) centered data matrix.
Consider the singular-value decomposition
\[
    \mathbf Y
    =
    \mathbf U\mathbf S\mathbf V^\prime.
\]
Its rank-\(L\) approximation is
\[
    \mathbf Y_L
    =
    \mathbf U_L\mathbf S_L\mathbf V_L^\prime.
\]
One convenient normalization defines the estimated grid loading vectors as the columns of
\(\mathbf U_L\) and the corresponding score matrix as
\[
    \widetilde{\mathbf Z}
    =
    \mathbf S_L\mathbf V_L^\prime .
\]
The \(t\)-th column,
\(\widetilde{\mathbf z}_t\),
is then a low-dimensional representation estimated directly from the station data.
Equivalent normalizations lead to the same fitted common component
\(\mathbf Y_L\).

The resulting score vectors can be forecast by VAR, KNN, or another multivariate method.
This route is included as an alternative benchmark rather than as the autocovariance-based dynamic-score estimator of Appendix~\ref{app:fpca}.
To obtain a continuous forecast map, the estimated mean and loading vectors must still be represented as surfaces.
In our implementation, they are reconstructed over \(\mathcal D\) using the finite-element basis from Appendix~\ref{sec:spatial smoothing}.

\subsubsection{Functional Autoregressive Benchmark}
\label{sec:EstFAR}

The functional autoregressive model of order one is
\begin{equation}
\label{eq:FAR1}
    X_t-\mu
    =
    \rho(X_{t-1}-\mu)
    +
    \eta_t,
\end{equation}
where \(\rho:\mathcal H\rightarrow\mathcal H\) is a bounded linear operator and
\(\{\eta_t\}\) is a functional white-noise process.
Standard conditions ensuring a stationary solution are discussed in
\citet[Theorem 3.1]{Bosq2000}.

Define the covariance operator
\[
    C_0 f
    =
    E\!\left[
    (X_{t-1}-\mu)
    \langle X_{t-1}-\mu,f\rangle
    \right]
\]
and the lag-one cross-covariance operator
\[
    C_1 f
    =
    E\!\left[
    (X_t-\mu)
    \langle X_{t-1}-\mu,f\rangle
    \right].
\]
Model~\eqref{eq:FAR1} implies
\[
    C_1=\rho C_0.
\]
Formally, \(\rho=C_1C_0^{-1}\).
However, \(C_0^{-1}\) is unbounded because the eigenvalues of the compact covariance operator converge to zero.
Regularization is therefore required.

Let
\[
    C_0 f
    =
    \sum_{j=1}^{\infty}
    \xi_j\langle f,\varphi_j\rangle\varphi_j,
    \qquad
    \xi_1\geq\xi_2\geq\cdots>0.
\]
A spectral-cutoff regularization replaces \(C_0^{-1}\) by
\begin{equation*}
    C_{0,J}^{-1}f
    =
    \sum_{j=1}^{J}
    \xi_j^{-1}
    \langle f,\varphi_j\rangle
    \varphi_j.
\end{equation*}
With sample estimators
\[
    \widehat C_0 f
    =
    \frac{1}{T-1}
    \sum_{t=1}^{T-1}
    (X_t-\widehat\mu)
    \langle X_t-\widehat\mu,f\rangle
\]
and
\[
    \widehat C_1 f
    =
    \frac{1}{T-1}
    \sum_{t=1}^{T-1}
    (X_{t+1}-\widehat\mu)
    \langle X_t-\widehat\mu,f\rangle,
\]
the regularized estimator is
\[
    \widehat\rho_J
    =
    \widehat C_1\widehat C_{0,J}^{-1}.
\]
The one-step forecast is
\[
    \widehat X_{t+1|t}^{\,FAR}(s)
    =
    \widehat\mu(s)
    +
    \widehat\rho_J
    \bigl(X_t-\widehat\mu\bigr)(s),
\]
and multi-step forecasts are obtained recursively.
The truncation level \(J\) is selected within the training sample.

\section{Data and Preprocessing}
\label{app:data_preprocessing}
\setcounter{equation}{0} \renewcommand{\theequation}{B.\arabic{equation}}
\setcounter{figure}{0} \renewcommand{\thefigure}{B.\arabic{figure}}
\setcounter{table}{0} \renewcommand{\thetable}{B.\arabic{table}}

This appendix documents the data construction and preprocessing steps used in the empirical illustration.
The goal is to make the main text concise while keeping enough detail for replication.

\paragraph{Raw ozone data.}
The ozone data come from AirBase, the European air-quality database provided by the European Environment Agency.
The raw archive contains daily ozone measurements for Germany, measured in \(\mu\mathrm{g}/\mathrm{m}^3\), with records available from a large set of monitoring stations.
For the empirical illustration, we restrict attention to the year 2011 and retain the \(N=171\) stations that operated continuously and had complete daily observations throughout the 365-day sample period.
The raw files were parsed into a unified data set containing the daily ozone observations and the corresponding station information.
The key variables are
\[
    \texttt{date}, \quad
    \texttt{latitude}, \quad
    \texttt{longitude}, \quad
    \texttt{ozone}.
\]

\paragraph{Meteorological covariates.}
As meteorological covariates, we use the E-OBS gridded data set.
The variables used in the analysis are mean temperature, precipitation, solar radiation, and wind speed.
The data are available on a regular \(0.25^\circ\times0.25^\circ\) latitude--longitude grid, corresponding approximately to a \(25\mathrm{km}\times25\mathrm{km}\) resolution over Central Europe.
The daily gridded observations were processed so that each meteorological variable could be represented as a surface over the same geographical domain and same time period as the ozone data.
The dataset is publicly accessible at: \url{https://surfobs.climate.copernicus.eu/dataaccess/access_eobs.php}.

\paragraph{Coordinate projection and modelling domain.}
The station coordinates are originally reported in WGS84 latitude--longitude coordinates.
For surface reconstruction and distance-based computations, all coordinates are projected to the Gauss--Kr\"uger Zone 3 coordinate system (EPSG:31467) using the MATLAB function \texttt{projfwd}.
All interpolation, triangulation, smoothing, and forecast-evaluation steps are then carried out in projected coordinates measured in meters.

The geographical modelling domain is determined by the station locations and the national boundary of Germany.
We construct a Delaunay triangulation over the projected station locations.
This gives an initial triangular mesh whose vertices are the monitoring stations.
To avoid extrapolating into areas outside Germany, we retain only triangles for which at least 75\% of their area lies within the national boundary.
This rule keeps the domain supported by observed stations while avoiding artificial smoothing outside the intended geographical region.
A stricter 100\% rule was also checked in earlier versions and produced negligible quantitative differences, but the 75\% rule gives a visually clearer and more coherent domain.
See Figure~\ref{fig:stations+mesh}.
The final domain covers approximately 89\% of Germany's land area and is used consistently in all reconstruction and forecasting steps.

\paragraph{Deseasonalization.}
Ozone and the meteorological covariates display strong seasonal patterns.
To focus the short-run forecasting exercise on deviations from the seasonal component, we remove monthly means.
For each variable and each geographical location, the monthly seasonal component is computed as the average over all days belonging to the same calendar month.
The deseasonalized observation is then obtained by subtracting this monthly component from the original daily value.
The estimated seasonal components are stored separately so that forecasts of deseasonalized surfaces can be mapped back into forecasts of the original concentration scale when required.

The supporting repository provides the cleaned data in the \texttt{Data} folder.
These files include \texttt{OzoneCleaned.csv} and the corresponding cleaned files for precipitation, solar radiation, temperature, and wind speed.
The folder also contains MATLAB files with the seasonally adjusted data and the estimated seasonal components, together with the \texttt{GeoConstraints} files used to define the German boundary.

\begin{figure}[tp]
\centering
  \includegraphics[width=0.9\textwidth]{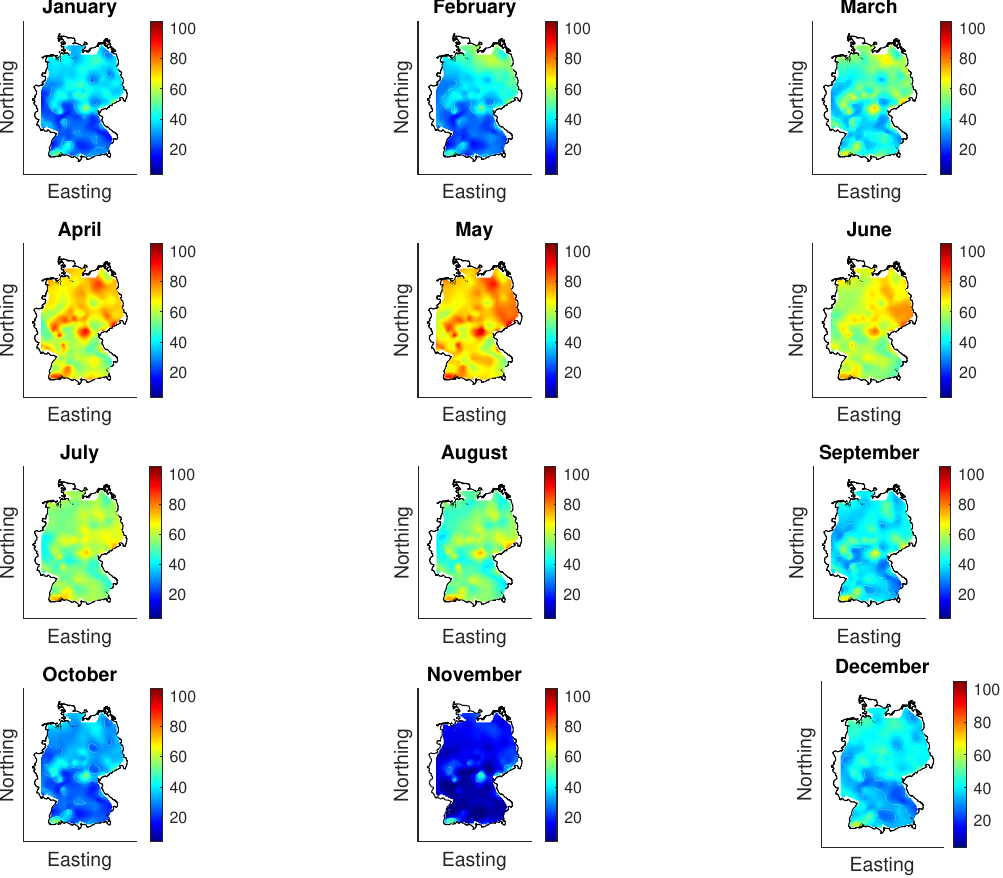}
  \caption{{\footnotesize \textbf{Monthly seasonal ozone concentration surfaces.}
  Each panel reports the estimated monthly average ozone surface.}}
  \label{fig:ozoneSeasonal}
\end{figure}

Figure~\ref{fig:ozoneSeasonal} shows the monthly ozone seasonal surfaces.
The seasonal component is highest on average during spring and early summer, with April--June displaying the largest average ozone levels over Germany and July--August showing localized high values in the region around Dresden.

\paragraph{Construction of the surface series.}
After deseasonalization, we reconstruct 365 daily surfaces for ozone, temperature, solar radiation, wind speed, and precipitation using the finite-element procedure described in Appendix~\ref{sec:spatial smoothing}, with smoothing parameter $ \lambda= 1e6$.
The same modelling domain is used for all variables.
The resulting objects contain the finite-element coefficient matrices, the common basis and triangulation, the estimated monthly seasonal components, and the corresponding station or grid information.
These objects form the input to the dynamic-score, FPCA-score, and forecasting procedures described in Appendix~\ref{app:forecasting_methodology}.

\section{Additional empirical outputs}
\label{sec:AdditionalResults}
\setcounter{equation}{0} \renewcommand{\theequation}{C.\arabic{equation}}
\setcounter{figure}{0} \renewcommand{\thefigure}{C.\arabic{figure}}
\setcounter{table}{0} \renewcommand{\thetable}{C.\arabic{table}}

The main text reports the one-step-ahead results.
For completeness, this appendix presents the comparison of FPCA and dynamic directions and the corresponding boxplots and MSPE surfaces for the 3-step- and 7-step-ahead forecasts.

\begin{figure}[th]
    \centering
    \includegraphics[width=0.8\textwidth]{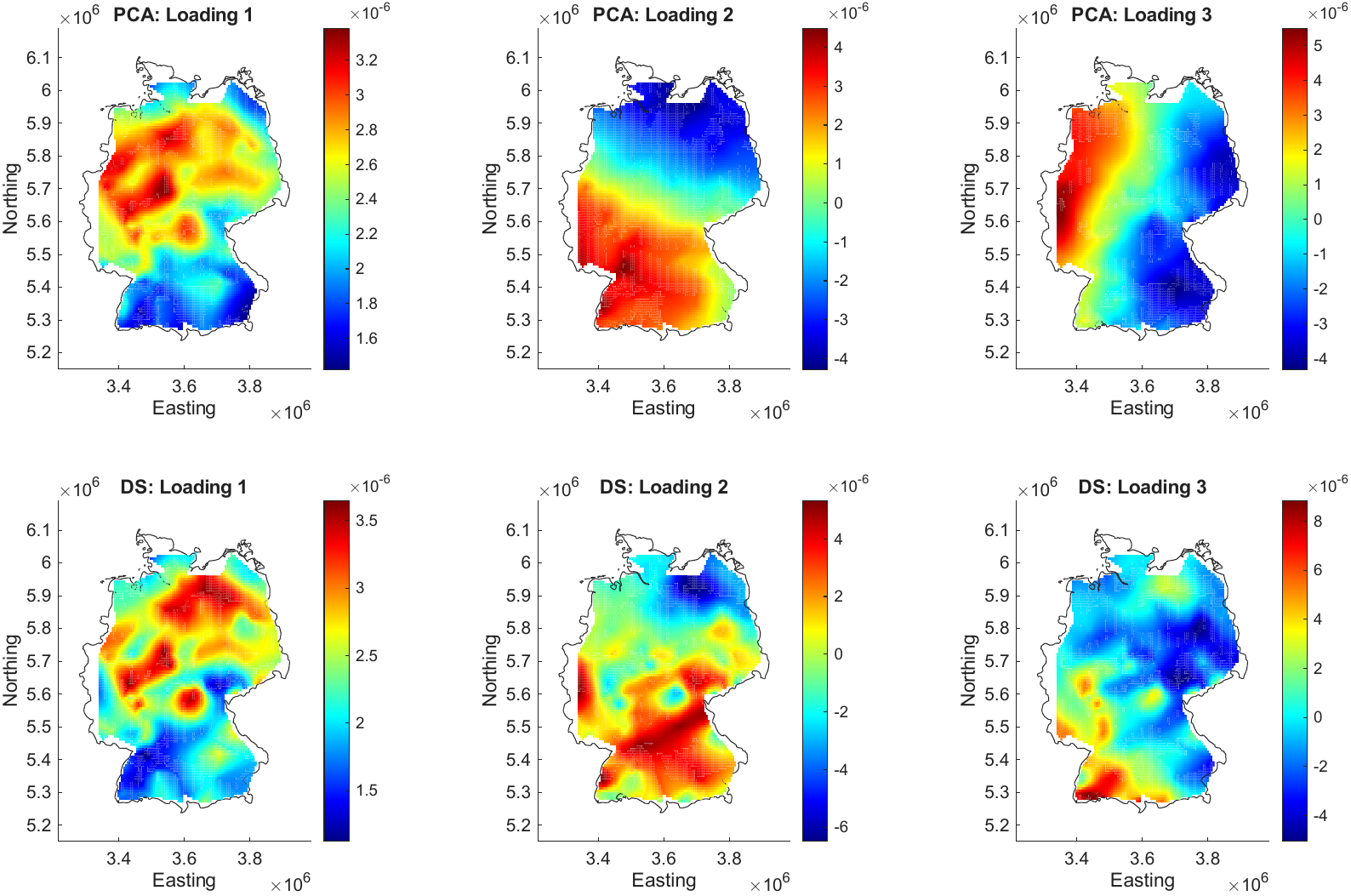}
    \caption{{\footnotesize \textbf{Leading FPCA and dynamic directions/loadings.} The figure compares the leading variance-based FPCA directions with the directions obtained from the cumulative autocovariance operator.}}
  \label{fig:PCA_DS_directions}
\end{figure}

\begin{figure}[tp]
\centering
  \includegraphics[width=0.8\textwidth]{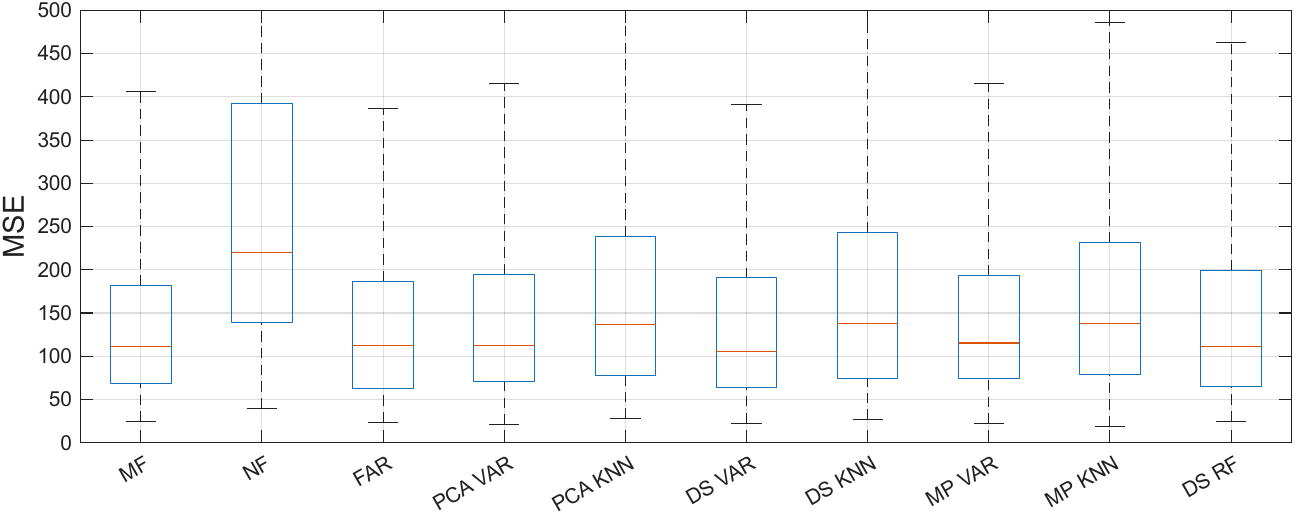}
  \caption{{\footnotesize \textbf{Boxplots of 3-step ahead forecast errors.}
  The figure reports rolling-origin forecast errors for the benchmark and surface-time-series methods.
  MF: mean forecast; NF: naive forecast; FAR: functional autoregression; PCA\_VAR/PCA\_KNN: FPCA-score forecasts; DS\_VAR/DS\_KNN: dynamic-score forecasts; MP\_VAR/MP\_KNN: multivariate-perspective forecasts; DS\_RF: random forest on dynamic scores.}}
  \label{fig:Forecasts3}
\end{figure}

\begin{figure}[tp]
\centering
  \includegraphics[width=0.8\textwidth]{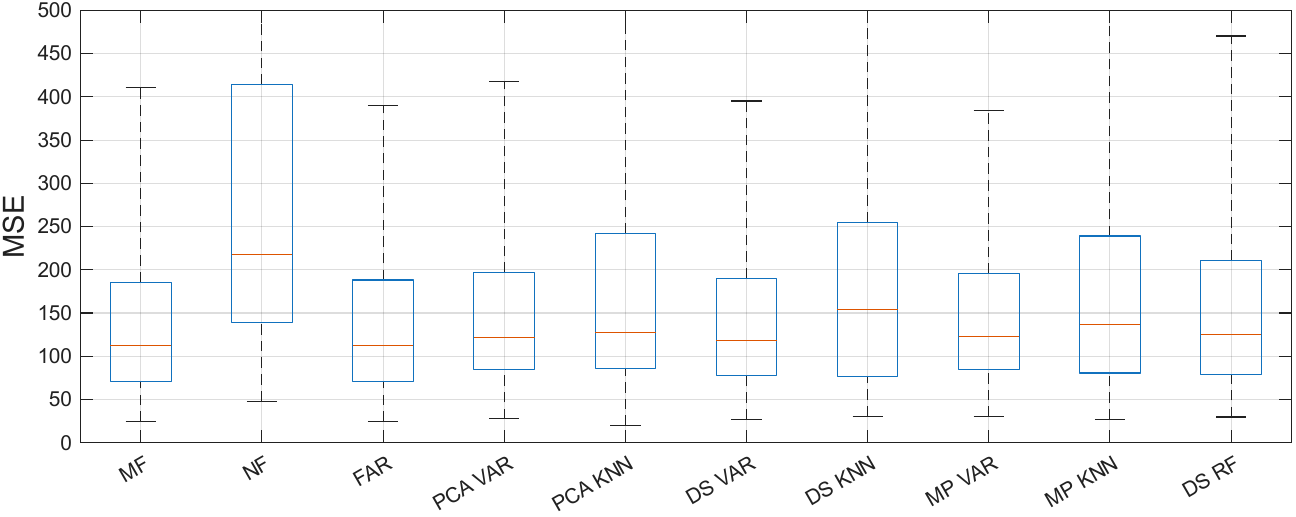}
  \caption{{\footnotesize \textbf{Boxplots of 3-step ahead forecast errors.}
  The figure reports rolling-origin forecast errors for the benchmark and surface-time-series methods.
  MF: mean forecast; NF: naive forecast; FAR: functional autoregression; PCA\_VAR/PCA\_KNN: FPCA-score forecasts; DS\_VAR/DS\_KNN: dynamic-score forecasts; MP\_VAR/MP\_KNN: multivariate-perspective forecasts; DS\_RF: random forest on dynamic scores.}}
  \label{fig:Forecasts7}
\end{figure}

\begin{figure}[tp]
\centering
  \includegraphics[width=0.85\textwidth]{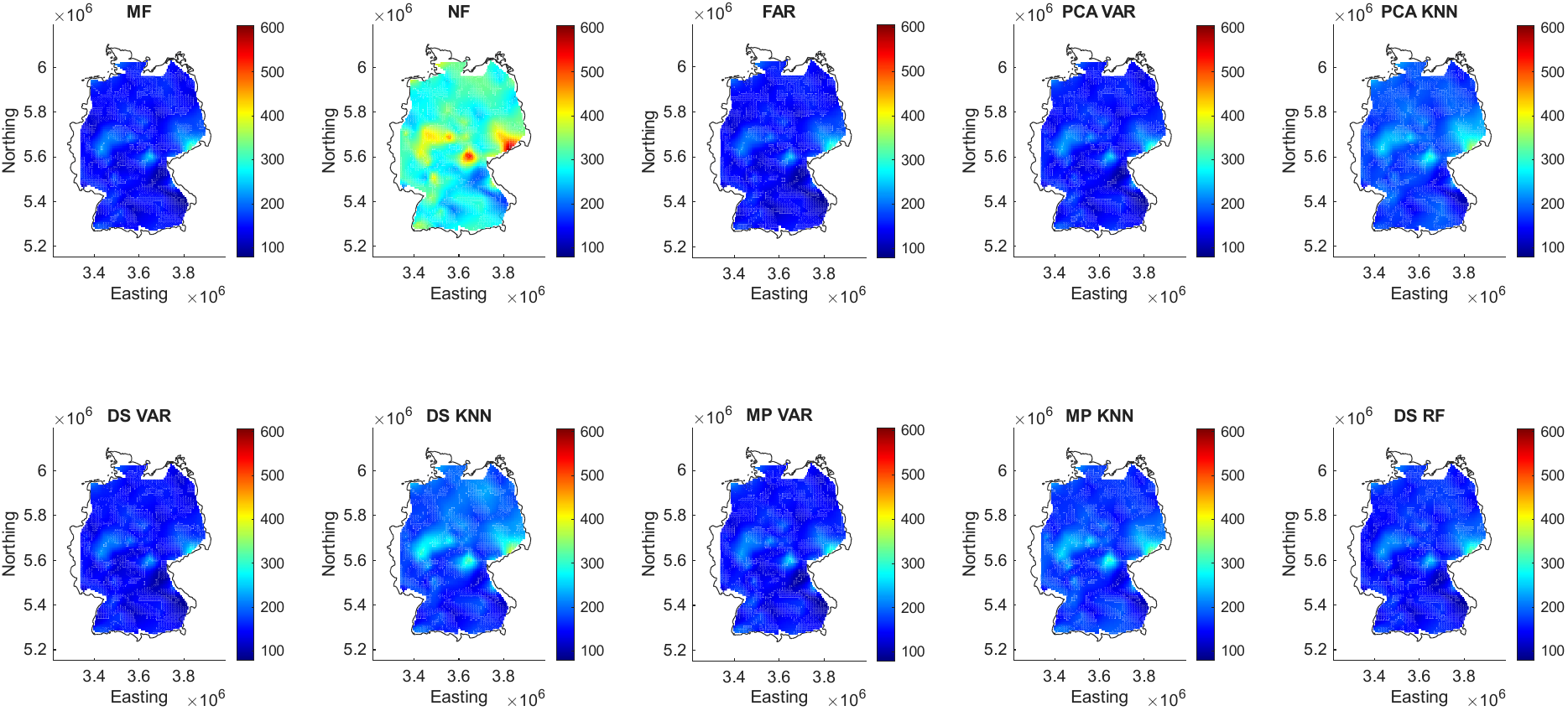}
  \caption{{\footnotesize \textbf{MSPE surfaces for 3-step-ahead forecasts.}
  Each panel shows \(\mathrm{MSPE}_{3}(s)\) over the reconstruction domain.
  The panels reveal where each forecasting method performs well or poorly across Germany.}}
  \label{fig:MSESurfaces3}
\end{figure}

\begin{figure}[tp]
\centering
  \includegraphics[width=0.85\textwidth]{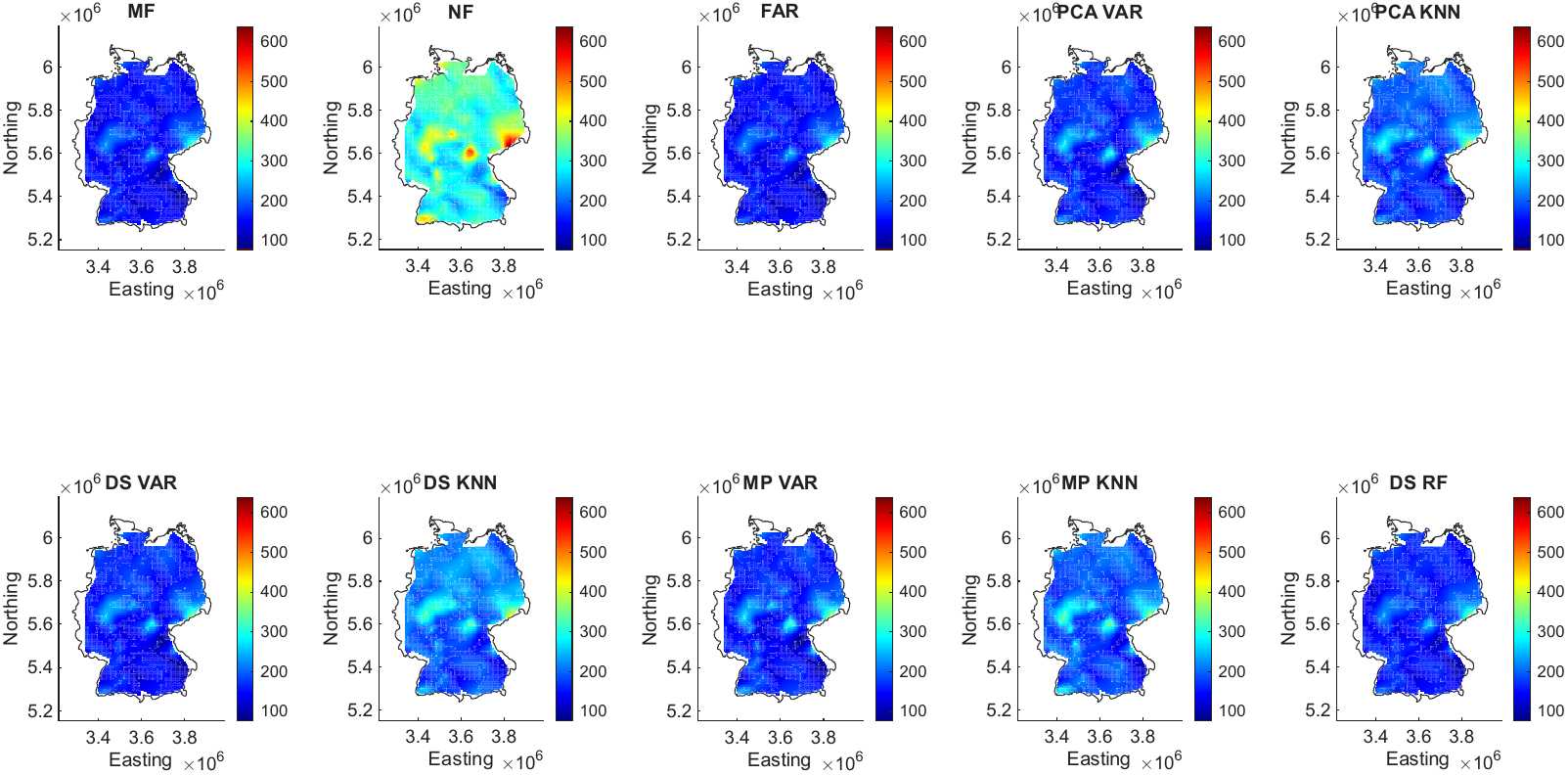}
  \caption{{\footnotesize \textbf{MSPE surfaces for 7-step-ahead forecasts.}
  Each panel shows \(\mathrm{MSPE}_{7}(s)\) over the reconstruction domain.
  The panels reveal where each forecasting method performs well or poorly across Germany.}}
  \label{fig:MSESurfaces7}
\end{figure}


\end{document}